# Evaluating the Impact of AI Standards: The use case of entity resolution

Julia Lane

Professor

New York University

Julia.lane@nyu.edu

# 1. Introduction to entity resolution

This paper is intended to provide an overview of how the evaluation of standards could be applied to entity resolution, or record linkage. Data quality is of critical importance for many AI applications, and the quality of data, particularly on individuals and businesses, depends critically, in turn, on the quality of the match of entities across different files.

Entity resolution is the technical term for ensuring that information about entities can be combined because the entities are the same – addressing the "problem of extracting, matching and resolving entity mentions in structured and unstructured data".[1] Getting entity resolution right is important, because high quality data on entities like people or organization are essential to many AI systems; creating high quality data increasingly requires correctly classifying information that comes from different sources as generated by the same entity. But it is also very difficult because data on the same entity that are acquired from different sources are often inconsistent and have to be carefully reconciled. The use of AI, in the form of machine learning methods, is becoming increasingly important because other approaches are less applicable for modern needs. In particular, manual methods to link data are too costly and slow to scale, and probabilistic methods are inappropriate in the increasingly frequent cases where unique identifiers are not available.[2]

There are at least three additional reasons why entity resolution is an important use case for evaluating the impact of AI standards. Entity resolution is an important building block in many AI systems. The impact of AI standards on current record linkage practices is likely to be measurable and able to be evaluated.

First, entity resolution is a common component of AI systems. Entity resolution is a building block technology that is both central to achieving many of the goals of AI systems and often associated with many of the harms. The reason is that few AI systems are based on data that are ideally fit for purpose; in most sectors, data have to be linked from a variety of different sources. In Information technology — Artificial intelligence (AI) — Use cases ISO/IEC TR 24030 – 2021 for example, many of the 132 use cases[1] described rely on data that have been linked, appended or joined from heterogeneous sources, or that need duplicate data elements resolved.

Second, AI standards for entity resolution are consistent with US government and NIST priorities. Partnerships across the private, academic, and public entities, which are common to addressing specific entity resolution problems, is a goal of the United States Government National Standards Strategy for Critical and Emerging Technology [3]. Data and knowledge have been identified by NIST as one of nine areas of focus for AI standards [4]. In addition, the input of AI actors in different stages of entity resolution is necessary to both identify and realize the opportunities offered from data linkage and to mitigate the harm that could result from incorrect links, from unethical use, or from confidentiality breaches. Thus the development of AI standards for entity resolution could inform the implementation of the NIST AI RMF trustworthiness requirements to other sectors. The subset of examples selected in the following sections include references to such general concepts as (i) terminology and taxonomy[2] (ii) Measurement and mitigations for risks and safety issues[3] (iii) Testing, evaluation, verification, and validation (TEVV) (iii) Risk based management of AI systems[4] (iv) security[5] and (v) Transparency among AI actors about system and data characteristics. [5].

[1] These use cases cover domains ranging from agriculture to home/service robotics, from construction to media, and from legal to security, defense and energy and deployments from cloud services and cyber-physical systems; from embedded systems to social networks

Finally, the impact of AI standards in providing implementation guidance for entity resolution is likely to be measurable relative to a counterfactual. Record linkage has been extensively used to combine datasets for at least three decades[6, 7], and is an important building block for many empirical problems, AI tools like machine learning have largely not been adopted at scale[6]. Thus the results of using rule based or probabilistic matching are well known, and the impact of adoption of AI standards can be compared to those results. Because AI standards are just now being developed, and are likely to be adopted at different time periods, in different context, and in different sectors, it should be possible to construct counterfactuals.

Standards for entity resolution could be particularly helpful in the use case of linking learning records to employment records. Section 2 discusses how the key AI actors can be identified both for the specific Learning Employment Record (LER) use case. Section 3 describes how they might provide input into both the theory of change and the development of AI standards for specific components of the LER use case. The concluding section and the appendix describe how the lessons that might be learned from this use case could be applied to different sectors (horizontally) or for different AI technologies (vertically).

---

[2] Building on Information technology — Artificial intelligence — Artificial intelligence concepts and terminology ISO/IEC 23 22989:2022

[3] Building on Towards a Standard for Identifying and Managing Bias in Artificial Intelligence NIST SP 1270

[4] Building on NIST AI RMF and Guidance on Risk Management ISO/IEC 23894:2023

[5] Building on Adversarial Machine Learning: A Taxonomy and Terminology of Attacks and Mitigations NIST AI 100-2 E2023

[6] More generally, as Vollmer et al. note with health care data, "Machine learning, artificial intelligence, and other modern statistical methods are providing new opportunities to operationalise previously untapped and rapidly growing sources of data for patient benefit…the literature as a whole lacks transparency, clear reporting to facilitate replicability, exploration for potential ethical concerns, and clear demonstrations of effectiveness".8. Vollmer, S., et al., *Machine learning and artificial intelligence research for patient benefit: 20 critical questions on transparency, replicability, ethics, and effectiveness.* bmj, 2020. **368**..

## 2. Identify key AI actors

AI systems can be seen as socio-technical in nature. Societal input, in the form of key AI actors, is critical to determining the degree to which the characteristics of trustworthy AI systems - validity and reliability, safety, security and resiliency, accountability and transparency, explainability and interpretability, privacy, fairness with mitigation of harmful – are met. Because the adoption of AI standards is largely voluntary, key AI actors are also critical to the success of AI standards: the effectiveness of AI standards depends on the degree to which the standards are implementable and useful[9].

In other words, the role of key AI actors is to provide context, information about sector norms, technical, societal, legal and ethical boundaries, identify tradeoffs that need to be made in AI standards development, and ensure that the AI standards are useful in addition to being technically sound

Many of the roles and tasks of key AI actors at all stages of the AI lifecycle are identified in the AI RMF (Figure 3 and Appendix A in the AI RMF). It is worth noting that the population from which key AI actors can be drawn is unlikely to be static. If the AI standards are successful, the population of key AI actors will grow, if they are unsuccessful, it will decline.

In the entity resolution use case, there are a number of ISO standards. These include improving data quality[7], measuring machine learning performance[8], managing risks[9], and reducing bias[10]. Key AI actors can be identified in industry, academia, government, and philanthropic foundations to provide a sociotechnical lens on many of these standards.

[7] ISO-IEC JTC 1-SC-42-WG2
[8] Information Technology — Artificial Intelligence — Assessment Of Machine Learning Classification Performance (ISO/IEC TS 4213:2022, IDT) specifies methodologies for measuring classification performance of machine learning models, systems and algorithms.
[9] SO/IEC 23894: 2023 Information Technology - Artificial Intelligence - Guidance On Risk Management provides guidance on how organizations that develop, produce, deploy or use products, systems and services that utilize artificial intelligence (AI) can manage risk specifically related to AI and how to integrate risk management into their AI-related activities and functions.

As businesses pivot, in an era with rapid technological change, to hiring employees based on skills and credentials, LERs have become increasingly important [10]. The US Chamber of Commerce has established JEDx[11] a "public-private approach for collecting and using standards-based jobs and employment data that will enhance government reporting, workforce analytics, and empower people to use their own records to pursue opportunities and advancement".

Beyond the specific case of LERs, many consulting firms like Accenture, Amazon, or Deloitte offer entity resolution as part of their data science services[12]. Some firms, like Senzing [11], have built an entire business of entity resolution as a service; others, like Thomson Reuters, have developed a separate product line (CLEAR) targeting other business and government in providing online investigation, fraud prevention and risk detection tools through proprietary record linkage of public data[13]. Business uses range from manufacturing (which requires entity resolution to match parts from different suppliers), to health services (which links electronic health records from different health care providers), and to advertising (which targets potential consumers).

In academia, there is a variety of communities. The breadth of literature makes it impossible to summarize; Binette and Steorts list statistics, computer science, machine learning, political and social science, medicine and epidemiology, official statistics, human rights statistics, author name disambiguation, and forensic science [2]. There are many professional societies, including the International Population Data Network which has almost 2,000 members from over 40 countries, and hosts biannual conferences[14].

Government agencies are also starting to build communities of practice around record linkage. For example, a community of state workforce and education agencies - has formed regional collaboratives that are linking those records and developing common standards[12, 13]. They are supported by the federal Department of Labor's Employment and Training Administration (ETA) and philanthropic foundations.[13, 14] National

---

[10] ISO/IEC TR 24368:2022 Information Technology - Artificial Intelligence - Overview Of Ethical And Societal Concerns provides information in relation to principles, processes and methods in this area; is intended for technologists, regulators, interest groups, and society at large.

[11] https://www.uschamberfoundation.org/solutions/workforce-development-and-training/jedx

[12] See, for example, Amazon's offering https://aws.amazon.com/entity-resolution/ and Accenture's acquisition https://newsroom.accenture.com/news/2018/accenture-forms-strategic-alliance-and-invests-in-data-analytics-firm-quantexa

[13] https://legal.thomsonreuters.com/en/products/clear

[14] https://ipdln.org/about-ipdln/

Association of State Workforce Agencies as the administrative organization for the state regional collaboratives[15], which hosts annual meetings as well as training programs.

Philanthropic foundations like the Gates Foundation have also invested substantial portions of their portfolio in supporting record linkage services like that provided by the National Student Clearinghouse. Walmart and the Charles Koch Foundation are funding LERs as part of a skills-based hiring ecosystem[16].

## 2.1 AI actors in the planning and design phase

AI actors in the first part of the entity resolution lifecycle are those who define the goals of the entity resolution process.  Simply put, the AI actors in this phase of the process are defined as those who determine how AI standards will be useful and how they will be implemented.

The AI standards that apply to their tasks would be  (i) identifying the need for entity resolution, (ii) mapping the need to business objectives and metrics, and (iii) defining the associated business requirements ( Stage 1 and Stage 2 in Information technology — Artificial intelligence — Data life cycle framework ISO/IEC 8183-2023) and  also (iv) do a risk and opportunity assessment (6.1 Information technology — Artificial intelligence — Management system ISO/IEC 42001-2023).

Identifying the need for high quality data in the LER use case is typically the role of senior managers in businesses – particularly their human resources departments, education agencies, workforce agencies, and education providers in the private and public sector, as well as for profit companies.  Senior agency managers typically require high quality information that can be implemented and used so that they hire the right employees - that firms have better workers in order to innovate and grow - educational institutions know what skills to provide to provide, so that workers and students have better workforce outcomes, and so that the cost to the taxpayer is not too great. Senior managers in the private sector could be producing detailed reports on workforce needs in particular sectors[17], leading technology companies that need to have better information about the workforce skills in different local labor markets, institutions of higher education and companies that provide alternative pathways for technical training  that need more information about what types of skills to provide[15].   The AI standards could provide a guiding structure for decision making.   AI standards could also be developed to help senior managers achieve additional goals such as protecting the confidentiality of individual level data, and minimizing bias in the analytical results.

[15] https://www.naswa.org/partnerships/multi-state-data-collaboratives/multi-state-data-collaboratives
[16] https://charleskochfoundation.org/stories/learning-and-employment-records-are-revolutionizing-hiring
[17] https://blogs.microsoft.com/blog/2024/05/08/microsoft-and-linkedin-release-the-2024-work-trend-index-on-the-state-of-ai-at-work/ ; https://www.reveliolabs.com/

The business objectives AI standards could guide which datasets get combined, and how the data are structured. For example, if the business objective is to identify the reasons why individuals who start training programs do not complete, a key source dataset would be the initial enrollment files of all students or trainees. Alternatively, if the business objective is to find out the job placements of individuals with particular certification, a key source dataset would be graduation files, combined with transcript information. In another example, if the business objective was to identify firms that are the largest employers of both completing and non-completing students, a key source dataset would be workforce data aggregated to the firm level, linked to job posting data and linked to enrollment files. The business objective could be developed in conjunction with those likely to use the results, particularly students, trainees, educational institutions, and employers.

The AI standards associated with the business requirement for entity resolution could also include a risk and opportunity assessment of different data sources. The key AI actors could use the AI standards to assess whether surveys are able to provide the longitudinal information on student educational and employment trajectories that is necessary to understand the supply of and demand for the trained workforce in a cost effective manner. The AI standards could provide a structure to identify the implications of choosing surveys on the ability to examine outcomes for under represented minorities. The key AI actors could also develop AI standards to asses the costs and benefits of the use of data records generated from the administration of education and workforce programs.[16], as well as from job postings[17, 18].

Finally, the AI standards could provide guidance to assess the confidentiality risks with the entity resolution necessary to link data across these multiple data sources [14, 19].

## 2.2 AI actors in the data and input stage

The AI actors in the second part of the entity resolution lifecycle are those who are responsible for producing the data required to address the business goals. These key actors will largely be data scientists, data engineers, domain exports, and TEVV experts. They could include private sector providers of entity resolution services and software. In the case of linking education and workforce data, many agency staff have neither the time nor the capacity to do the actual linkages. As a result, private sector providers often take the lead in developing the data framework, including the metadata standards (the terminology and taxonomy), as well as the measurements of the concepts. They could also include human resources department of businesses, government agency staff, training providers, and institutions of higher education who understand how the data have been generated.

The key AI actors will need to provide input on the development of AI standards in a number of areas. These include the methods that should be used to standardize the different data files as identified in Building on Information technology — Artificial intelligence — Artificial intelligence concepts and terminology ISO/IEC 23 22989:2022. It could also include documenting the impact of different decisions on different social groups - the terminology and taxonomy, since even the evaluation of demographic impact can vary systematically across entities that differ on the categorization of black, white, Hispanic and historically excluded sub-populations[18].

Their tasks consists of identifying standards that would inform the following (i) planning the scope of the data collection , (ii) acquiring the data, and (iii) preparing the data ( Stage 3, 4 and 5 in Information technology — Artificial intelligence — Data life cycle framework ISO/IEC 8183-2023) (iv) documenting the work and (v) communicating any data issues to senior management (Section 6 Information technology — Artificial intelligence — Management system ISO/IEC 42001-2023).

### 2.2.1 AI actors and standards in the scope of data collection

AI actors could help design standards for linkage. The canonical linkage task is described in Figure 1. For simplicity's sake, assume that there are two files: A and B. The data scientist has to look at each row in data file A and compare it to each row in data file B. Then she has to decide whether or not the two rows reference the same person. Data might be linked based on common identifiers like social security numbers, but often those data are insufficient (like education data for low income and highly mobile populations), noisy, or not available. Often problems are even harder, if they require matching across files which did not have the same identifiers – employment records might have SSN and name, which do not link easily to education records which have an institution specific number and name. So data have to be linked by name, geographic location, by patterns of behavior, or other identifying information. It would be extraordinarily expensive, if not impossible, for an analyst to do such a match manually because of the scale issues. If there are 25 rows in each file, that would mean that the analyst would have to make 625 pairwise comparisons. If there are 100 in each file, then there are 10,000 comparisons, putting it out of the reach

[18] See, for example, Appendix G: Classification Of Demographics, Race-Ethnicity-URM, 2023 MULTI-STATE POSTSECONDARY REPORT. https://kystats.ky.gov/Latest/MSPSR

of human processing. A million rows in each file generate a trillion possible pairwise comparisons.

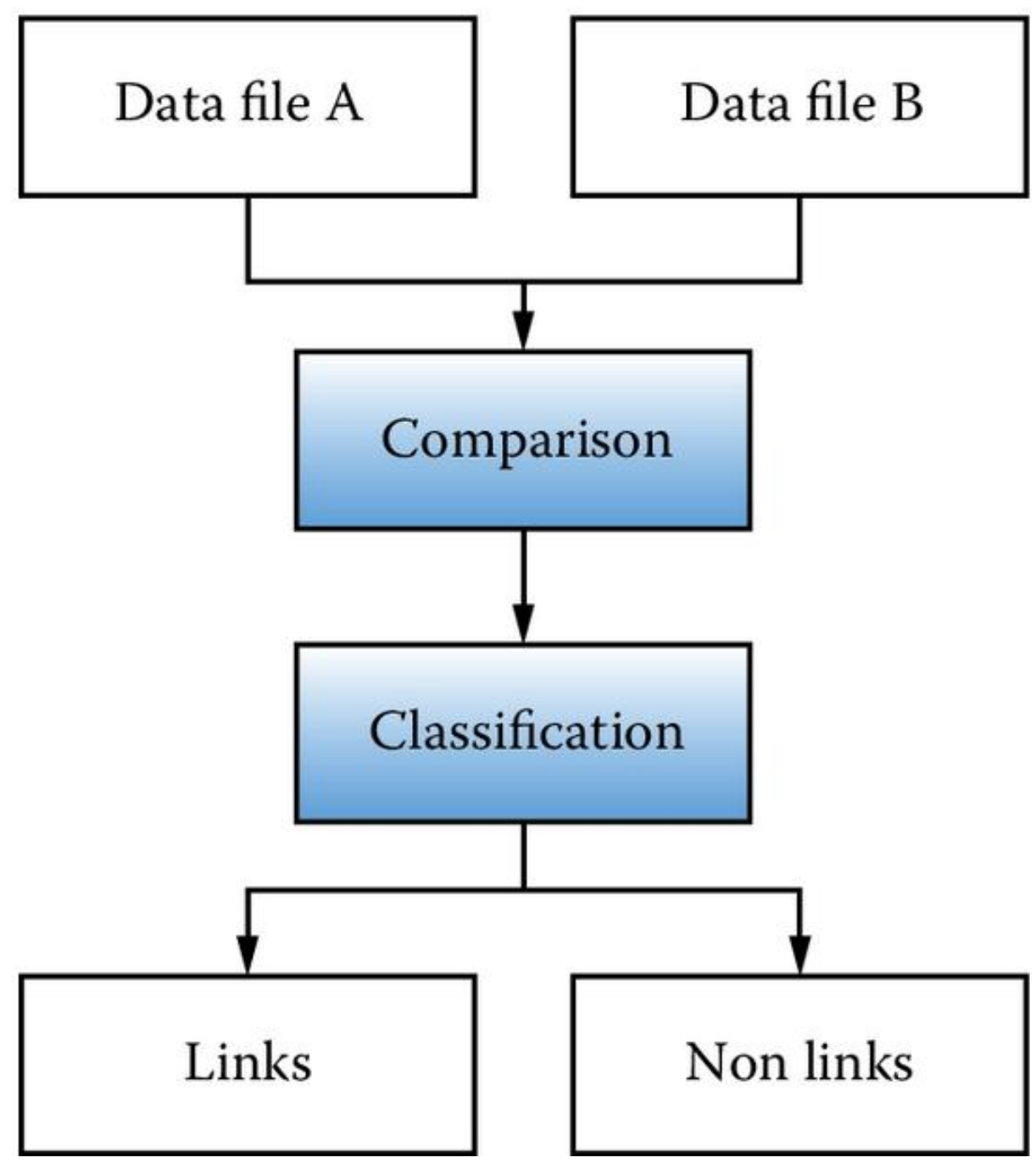

*Figure 1:The basic linkage problem from (Fig 3.1) Big Data and Social Science https://textbook.coleridgeinitiative.org/*

### 2.2.2 AI actors and standards for data acquisition

The confidential nature of both learning and employment records means that another set of key AI actors - lawyers and data owners – are likely to need to be involved to reduce the risk of legal liability. This is consistent with Building on NIST AI RMF and Guidance on Risk Management ISO/IEC 23894:2023 and repeated call for legal standards in general and AI standards in particular that provided common templates and language for record linkage.

### 2.2.3 AI actors and standards for preparing the data

The validity and reliability of machine learning models for record linkage often depends on the quality of the data on which they are trained. While the Data Lifecycle Framework provides important guidelines (ISO/IEC 8183) for standardization, their adoption in the field of entity resolution has been limited[19].

---

[19] Their specific data preparation steps include Cleaning; Feature engineering; Normalizing and scaling; Data organization; Labelling; Enrichment: De-identification.

Failure to link learning and employment records correctly will lead to poor quality LERs. The key AI actors could provide input on the development of AI standards in a number of areas.

One area would include the methods that should be used to standardize the different data files and the impact of different decisions on different social groups including the metadata standards (the terminology and taxonomy), as well as the measurements of the concepts.

AI standards are likely to need to be quite specific. The key AI actors trying to use data from different sources might have to deal with data that come in different formats, have missing or incorrect values, or contain inconsistencies. The approach could go beyond tools and institute processes as well; the high level requirements are identified in ISO/IEC 22989:2022). Preprocessing helps address these issues, ensuring that the data is in the best possible shape prior to linkage. This step is crucial because the quality of the preprocessing work can significantly impact the accuracy and usefulness of the linked data. Poor preprocessing might lead to errors or biases in the training and test datasets [6].

Another area might be incorporating understanding of the data generating process[20, 21].[20] In the LER use case, missing data might not result from model error, for example. It could be generated by people leaving the program or leaving the labor market for family reasons, and hence should not be included in a training dataset for classifying employment outcomes.

Academic researchers who are experts in data linkage and data analysis are also important here. They could help inform the development of AI standards to, for example, provide common look up tables across different datasets. This could include obvious factors, like making sure that missing value codes are the same in both fields[22]. They could also help develop common metadata standards, with substantial effects on the ultimate business need. In a fairly well known example of the effects of state metadata differences, for example, "the definitions for death or hospitalisations due to COVID-19 infections, where different US states used various definitions that resulted in databases that could not be used for comparative analysis. Unless metadata are available that clearly describe such definitions and their changes, it can be difficult to identify the effects of any changed definitions because any such change might have subtle effects on the characteristics of only some individuals in the population of interest for a research study".[7].

[20] https://www.immport.org/shared/home; https://globalbiodata.org/what-we-do/global-core-biodata-resources/

*Representatives* of civil society likely to be harmed by erroneous linkages can also inform the development of AI Standards by providing information about how to measure and mitigate risks and safety issues. Their input could help address the impact of low quality linkages on minority populations. For example, because of the sheer size of the files, it is common to block by some common identifiers, like residential zip code or first letter of last name, and linkage quality may be degraded for mobile low income populations or women who marry and change names.

### 2.2.4 AI actors and standards for secure data linkage

Key AI actors for this component could include cybersecurity experts who can establish the initial infrastructure of the linkage system to reduce the risks of harm to people, organizations, and the ecosystem [5]. One important area is to reduce individual harm by reducing the likelihood of individual data being identified is AI standards that ensure that linkage take place in *secure, safe, and resilient environment*. This also reduces the risk of harm to organizations, like federal and state government agencies, who are statutorily required to protect data on participants in government programs or on survey respondents and which can be negatively affected by data breaches. Providing access is critical, however, to ensure that the utility of the data linkages is not corrupted by excessively restrictive decisions.[23]

Key AI actors in this case might be certified external assessment organizations. The FedRAMP program provides an example of how AI standards in the form of a standardized checklist can be used to ensure that approved practices (like identity management, continuous monitoring, encryption, and training) are adhered to by data users and that agencies can have their own cloud based secure environment with established access boundaries, so that no data can come into or out of the environment without approval. FedRAMP also has an infrastructure of certified external assessment organizations to verify that appropriate security is in place. There are over 400 different controls that are monitored by those assessors, including user training, audit logs, identity management, and incident response protocols.

Key AI actors, such as lawyers, academics, cybersecurity experts, and representatives of civil society might help inform the nature of AI standards associated with different access tiers. Access tiers are a combination of IT systems, services, and security controls appropriate for the sensitivity level of the data it hosts and protects. Tiered access is an application of data minimization, a key data protection safeguard for evidence building as embodied in the Fair Information Practice Principles. Information Security standards and practices support managing the risks to safety of people inherent to the record linkage use case described in this document. Data and system security broadly mitigate harms that can come from improper use, unauthorized disclosure, and theft of data resulting from the record and data element sharing necessary for record linkage. In particular, the NIST Cybersecurity Framework (CSF)[21], Cybersecurity and Infrastructure Security Agency Zero Trust Maturity Model 2.0[22], and the Federal Risk and Authorization Management Program[23] (FedRAMP) describe relevant approaches and actions to mitigate risks to the NIST AI RMF Safe and Secure and Resilient trustworthy characteristics.

Another could be to develop common Testing, Evaluation, Verification, and Validation (TEVV) practices in the development of the AI systems for the use case.

### 2.2.5 Communicating results back to all AI actors

Communication is essential to ensure there is transparency among AI actors about system and data characteristics Standardization could be seen as an ongoing process, where initial standards are not designed to meet all needs from the beginning, but can be adapted and changed. As noted in the AI RMF, it is important to include test, evaluation, verification, and validation (TEVV) processes throughout an AI lifecycle so that the operational context of an AI system can be generalized. .

## 2.3 AI Actors and Evaluating Entity Resolution Models

Key AI actors in this case would include individuals who could provide input into the contextual and technical model decisions to be included in the relevant AI Standards. These include individuals who choose the machine learning classification models, determine the thresholds, decide on the precision/recall tradeoff or other ML performance metrics. The actors would include:

*Government agency staff*, *training providers*, and *institutions of higher education* who can define the loss function. They will need to agree on the criteria used to measure how close a given row (a) in File A is to a row (b) in File B. In the case of different machine learning models, each will generate a score that is a measure of "closeness". Different thresholds will yield different tradeoffs with different metrics of success; two common ones are precision (ensuring that any specific link that is asserted to be correct is correct) and recall (ensuring that all possible correct links are captured). The costs and benefits to institutions, workers, firms, as well as to taxpayers would need to be assessed in the context of the use case, as well as potential biases [24].

[21] https://nvlpubs.nist.gov/nistpubs/CSWP/NIST.CSWP.29.pdf
[22] https://www.cisa.gov/sites/default/files/2023-04/zero_trust_maturity_model_v2_508.pdf
[23] https://fedramp.gov

*Academic researchers* and *industry representatives* who are up to date on the appropriate science, technology, and ethics issues in developing models would include computer scientists, social scientists, ethicists, as well as domain experts in education and workforce issues. In the case of AI standards, their expertise would be particularly important in terms of comparing entity resolution algorithms[25]

*Representatives* of civil society who can provide input into the loss function as well as verify and validate the output. This is particular important since *standard fairness criteria* for AI standards, such as outcome prediction instruments, cannot all be simultaneously satisfied when outcome prevalence differs across groups[26]. While the basic tool used to describe the tradeoff is the confusion matrix, many analysts in general are unfamiliar with the effect of different thresholds on the loss associated with incorrectly identifying two rows as a match (the precision) versus the loss associated with missing individuals who should have been matched, but who weren't (the recall). This is particularly important since there is a direct tradeoff between the two measures, as the threshold gets varied. Each cell of the matrix—true positives, true negatives, false positives, and false negatives—provides nuanced insights into the model's strengths and shortcomings. In some scenarios, the consequences of false positives might outweigh those of false negatives. In others, the opposite might hold true. Different groups are likely to have different views as to what that loss function should looks like. [27].

## 2.4 AI Actors in Reviewing Tasks and Outputs

As noted in the NIST AI RMF, trustworthiness in AI systems depends on there being transparency among AI actors about system and data characteristics. AI actors for this stage of the lifecycle will thus need to be able to provide input into how AI standards can provide that transparency. Those AI standards might include templates on how to structure analysis of the model output and apply bias audit toolkits [28], and require the input of trained data scientists, computer scientists, and social scientists from academia, government, and industry. Those same AI actors could inform the development of AI Standards reports about how the systematic biases by gender and ethnicity, which can occur with name-based matches, have been addressed. Similarly, they could inform the development of AI standards reports of the systematic biases that can occur if geographic data such as state or zip code are used to match records, and low income populations, including minorities, are known to move more than others[29].

Key AI actors could also provide input into how AI standards could report on the impact of privacy or confidentiality restrictions on the utility of the analysis. For example, restrictions on access can mean that it is difficult for communities to find out the magnitude of any biases, because analysts can't access the data--including meta data--to find out what errors have been made[16]. Those same AI actors could provide input into ways in which AI standards could report on the impact of attempts to protect confidentiality by means of, for example, synthesizing data or implementing differential privacy techniques, on the quality of the results and the ability of the AI system to achieve the stated goals. Such AI actors could include the research community, state agencies using the data, as well as industry [23]. In sum, the management of the risks of linking confidential data could be understood in the context of a transparently communicated and community defined value proposition.[5].

## 3. AI Actors and developing a theory of change

Once the key AI actors at each stage have been identified, the SDO and the actors need to unpack the theory of change associated with applying specific AI standards to the different parts of the record linkage process. This document uses an abbreviated theory of change [30]. The inputs, or resources at the disposal of the project, would be contributed both by members of the standards developing organization (SDO) and the key actors, in the form of time, workgroups, workshops, research papers. The activities and outputs generated by those inputs would consist of the joint development and production of the documentary AI standards. As noted in NIST AI 100-5, those AI standards would be designed to be implementable and useful. The critical part of the process would be the input of the key AI actors to the SDO to ensure that the target population would be able to implement and use the standards in the particular context of the entity resolution use case.

The identification of the elements of an abbreviated theory of change - the inputs, activities and outputs, outcomes and final goals –were identified in the AI standards index document. Examples of the application to the entity resolution use cases are sketched in Figure 2 in this document based on the discussion above. The key pieces of the theory of change are to describe the effect of the AI standards for entity resolution and outputs developed by the standards developing organization (column 2) on the outcomes (column 3), and final goals (column 4).

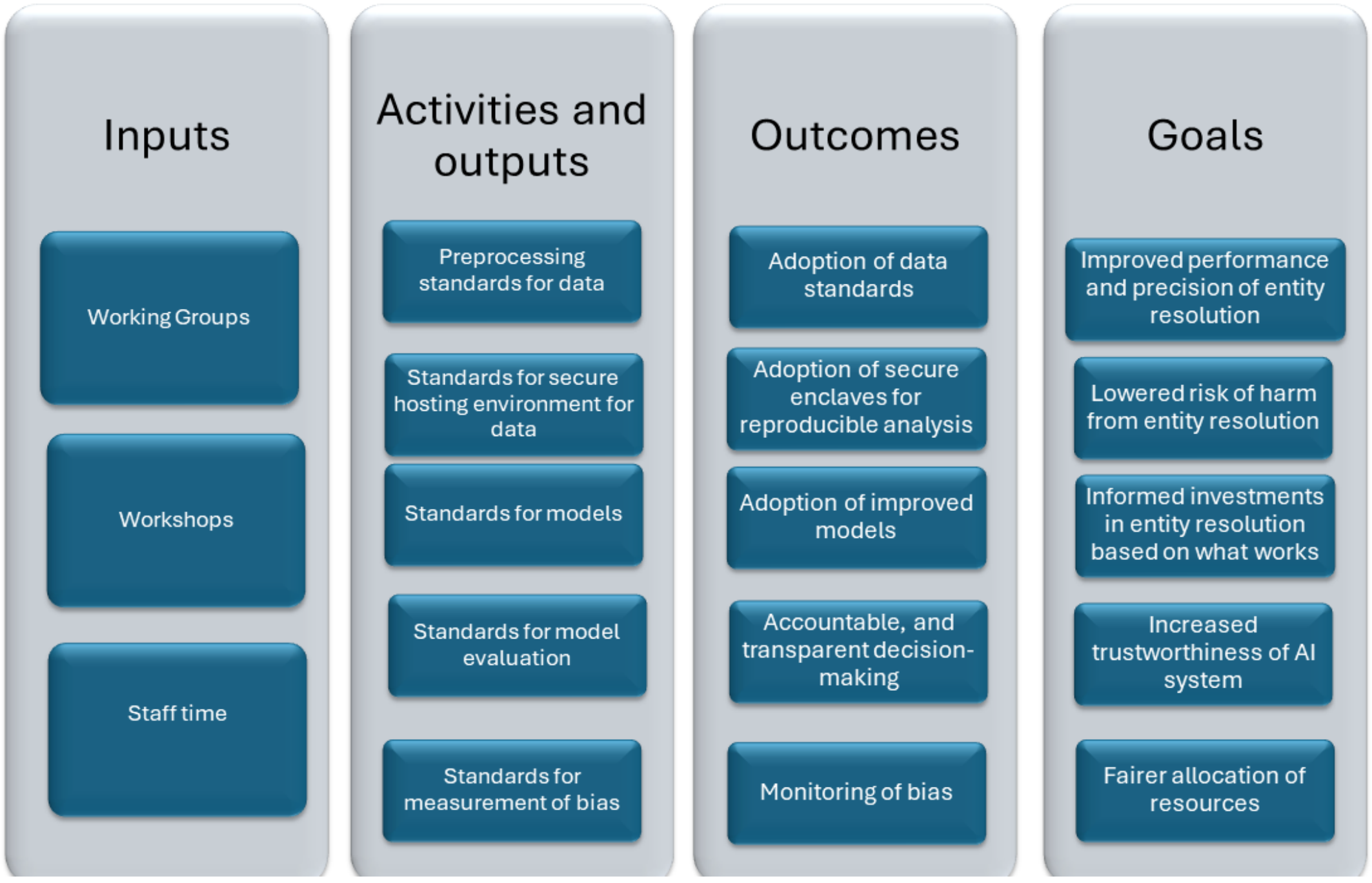

*Figure 2:An example of a theory of change for AI Standards in record linkage*

## 4. Concluding comments

This discussion of LERs as a use case is intended to be illustrative not exhaustive. It is just one of many in which AI standards could inform the development and adoption of AI record linkage tools. The use of record linkage and AI tools is not standardized and not always well understood. If AI standards led to greater adoption, the impact could range from improved forecasting, outlier detection and program management. The vertical applications are many – additional examples would include consulting, manufacturing, and health care.

In the case of government agencies, better entity resolution could result in the more efficient provision of permanent supportive housing, targeted investments in workforce training, the reduction in fraud due to collusion to the more effective detection of child abuse, or the effect of Supplemental Nutritional Assistance Programs on health outcomes. Other examples are provided in the supplemental materials of the Advisory Committee on Data for Evidence Building Year 2 report[24] [14]

[24] https://www.bea.gov/system/files/2022-10/supplemental-acdeb-year-2-report.pdf

# Appendix Examples

## Education

*The issue* The US Department of Education estimates that about $870 billion was spent on public elementary and secondary education in 2019-20[25] . Over $700 billion was spent in public, private and non for profit higher education institutions[26]. The US Department of Education notes that linking records is necessary to "improve classroom instruction, to measure student outcomes, and facilitate implementation of educational applications to evaluate the effectiveness of educational programs"[27] Linking records is necessary to ensure that Federal Student Aid is correctly disbursed[28]. The Advisory Committee on Data for Evidence Building noted that "Unprecedented changes in labor markets have led to fundamental changes in skill demands. Both sets of changes underscore the need to strengthen the connection between employment services, post-secondary programs, and workforce outcomes. More generally Governors, departments of labor, economic development planners, education and training providers, and unions can use better predictive information so they can plan for and support the growth of high wage jobs in their states. Students and parents have more information about career pathways.".[12]

*The role of AI standards* Linking confidential records must be done in a safe and secure manner (and is protected by the Family Educational Rights and Privacy Act (FERPA)[29]). Incorrect linkages would disproportionately affect marginalized communities, provide limited information about small demographic groups[31], fields of specialization, and geographic areas[32, 33].

*Key AI actors* State departments of education, as well as the federal National Center for Education Statistics that supports the Statewide Longitudinal Data Systems SLDS Grant program[30]. The SLDS program has annual meetings, a state data support team, and information about best practices. Affiliated organizations include institutions of higher education, particularly their professional association (State Higher Education Executive Officers Association[31]).

---

[25] https://nces.ed.gov/programs/digest/d22/tables/dt22_236.10.asp
[26] https://nces.ed.gov/programs/digest/d22/tables/dt22_334.10.asp ; https://nces.ed.gov/programs/digest/d22/tables/dt22_334.30.asp ; https://nces.ed.gov/programs/digest/d22/tables/dt22_334.50.asp
[27] https://studentprivacy.ed.gov/privacy-and-data-sharing
[28] https://fsapartners.ed.gov/knowledge-center/fsa-handbook/2023-2024/vol2/ch7-record-keeping-privacy-electronic-processes
[29] https://www2.ed.gov/policy/gen/guid/fpco/ferpa/index.html
[30] https://nces.ed.gov/Programs/SLDS/
[31] https://sheeo.org/

## Criminal Justice

*The issue* The US Department of Justice estimates that the monetary and social cost of crime is enormous. Arnold Ventures estimates that one in 14 US children have had an incarcerated parent, there are 2.2 million incarcerated adults, and the US spends over $80 billion annually on prisons, jail, probation and parole[32].

*The role of AI standards* There is "no unified data infrastructure for measuring the U.S. criminal justice system, evaluating its policies, or understanding the population that interacts with it"[34, 35]. There are thousands of different jurisdictions at the federal, state, local, and tribal level that capture data from courts, probation offices, prisons, jails, and parole offices, so that tracking individuals through the criminal justice system to provide services, as well as reduce the likelihood of reincarceration requires linking records. The challenges range from technological limitations and privacy concerns[33] to the lack of high quality identifiers to link confidential records[34, 35]. Incorrect linkages would disproportionately affect marginalized communities.

*Key AI actors* Federal, state, and local administrative entities, ranging from courts to county jails, to state prisons. The Federal Bureau of Justice Statistics State Justice Statistics Program[34] and Bureau of Justice Analysis University data platforms like the Criminal Justice Administrative Records System[35].

## Health:

*The issue* The US Department of Health and Human Services estimates that health care spending cost more than $4.5 trillion, or 17.3% of GDP in 2022[36] Health records[37]. Much of the health care costs are incurred by state run programs. "Medicare spending accounted for 21 percent of total national health care expenditures and reached $944.3 billion in 2022; Medicaid spending accounted for 18 percent of total health care expenditures, reaching $805.7 billion"[36].

*The role of AI standards* High quality identifiers are not typically available; ; linking confidential records must be done in a safe and secure manner (and is protected by the Health Insurance Portability and Accountability Act (HIPAA)[38]); incorrect linkages would disproportionately affect marginalized communities.

---

[32] https://www.arnoldventures.org/work/criminal-justice
[33] https://www.corrections1.com/probation-and-parole/predictive-analytics-identifying-risk-factors-and-targeting-resources-in-community-corrections
[34] https://bjs.ojp.gov/programs/state-justice-statistics-program
[35] https://cjars.org/
[36] https://www.cms.gov/data-research/statistics-trends-and-reports/national-health-expenditure-data/historical
[37] https://www.povertyactionlab.org/resource/acquiring-and-using-administrative-data-health-care-research
[38] https://www.hhs.gov/hipaa/index.html

## Food security:

*The Issue* The US Department of Agriculture spends over $100 billion a year on food stamps (Supplemental Nutrition Assistance Program), which benefits about 12.5% of the US population[39]. The program is managed by state agencies. Better linkages across agencies and states would improve the ability of state agencies to track program eligibility to ensure all beneficiaries are reached as well as minimizing fraud and evaluating program effectiveness [37].

*The role of AI standards* The High quality identifiers are not typically available; linking confidential records must be done in a safe and secure manner; incorrect linkages would disproportionately affect marginalized communities.

*Key AI actors* USDA Food and Nutrition Service (FNS) staff; Food stamp (or SNAP) administrators in each state; university public policy schools and schools of public health.

[39] https://www.pewresearch.org/short-reads/2023/07/19/what-the-data-says-about-food-stamps-in-the-u-s/